\documentclass[showpacs,amsmath,amssymb,twocolumn,prl,superscriptaddress]{revtex4-2}
\usepackage{amssymb}
\usepackage[dvips]{graphicx}
\usepackage{enumerate}
\usepackage{epsfig} 
\usepackage{subfigure}
\usepackage{xcolor}
\usepackage[T1]{fontenc}
\usepackage{fullpage}
\usepackage{amsthm,amsfonts,amssymb,amscd,mathrsfs,xspace,framed}
\usepackage{mathrsfs,amsmath}
\usepackage{color}
\usepackage{setspace}
\usepackage{url}
\usepackage{wrapfig}
\usepackage{tikz}
\usepackage{enumitem}
\usepackage{bm}
\usepackage{comment}
\usepackage{txfonts}
\usepackage{array}
\usepackage{color}
\usepackage{braket}
\usepackage{natbib}
\usepackage{cleveref}
\usepackage{graphicx}
\usepackage{pdfpages}

\def\be{\begin{equation}}
\def\ee{\end{equation}}
\def\ba{\begin{eqnarray}}
\def\ea{\end{eqnarray}}

\makeatletter
\AtBeginDocument{\let\LS@rot\@undefined}
\makeatother

\begin{document}

\title{Quantum Receiver Enhanced by Adaptive Learning}

\author{Chaohan Cui}
\affiliation{James C. Wyant College of Optical Sciences, The University of Arizona, Tucson, Arizona 85721, USA}
\author{William Horrocks}
\affiliation{James C. Wyant College of Optical Sciences, The University of Arizona, Tucson, Arizona 85721, USA}
\author{Shuhong Hao}
\affiliation{Department of Materials Science and Engineering, The University of Arizona, Tucson, Arizona 85721, USA}
\author{Saikat Guha}
\affiliation{James C. Wyant College of Optical Sciences, The University of Arizona, Tucson, Arizona 85721, USA}
\affiliation{Department of Electrical and Computer Engineering, The University of Arizona, Tucson, Arizona 85721, USA}
\author{\\N. Peyghambarian}
\affiliation{James C. Wyant College of Optical Sciences, The University of Arizona, Tucson, Arizona 85721, USA}
\affiliation{Department of Materials Science and Engineering, The University of Arizona, Tucson, Arizona 85721, USA}
\author{Quntao Zhuang}
\affiliation{Department of Electrical and Computer Engineering, The University of Arizona, Tucson, Arizona 85721, USA}
\affiliation{James C. Wyant College of Optical Sciences, The University of Arizona, Tucson, Arizona 85721, USA}
\author{Zheshen Zhang}
\email{zsz@arizona.edu}
\affiliation{Department of Materials Science and Engineering, The University of Arizona, Tucson, Arizona 85721, USA}
\affiliation{Department of Electrical and Computer Engineering, The University of Arizona, Tucson, Arizona 85721, USA}
\affiliation{James C. Wyant College of Optical Sciences, The University of Arizona, Tucson, Arizona 85721, USA}

\begin{abstract}
Quantum receivers aim to effectively navigate the vast quantum-state space to endow quantum information processing capabilities unmatched by classical receivers. To date, only a handful of quantum receivers have been constructed to tackle the problem of discriminating coherent states. Quantum receivers designed by analytical approaches, however, are incapable of effectively adapting to diverse environment conditions, resulting in their quickly diminishing performance as the operational complexities increase. Here, we present a general architecture, dubbed the quantum receiver enhanced by adaptive learning (QREAL), to adapt quantum receiver structures to diverse operational conditions. QREAL is experimentally implemented in a hardware platform with record-high efficiency. Combining the QREAL architecture and the experimental advances, the error rate is reduced up to 40\% over the standard quantum limit in two coherent-state encoding schemes.
\end{abstract}

\maketitle

\begin{figure*}[htb!]
\centering
\includegraphics[width=6.5in]{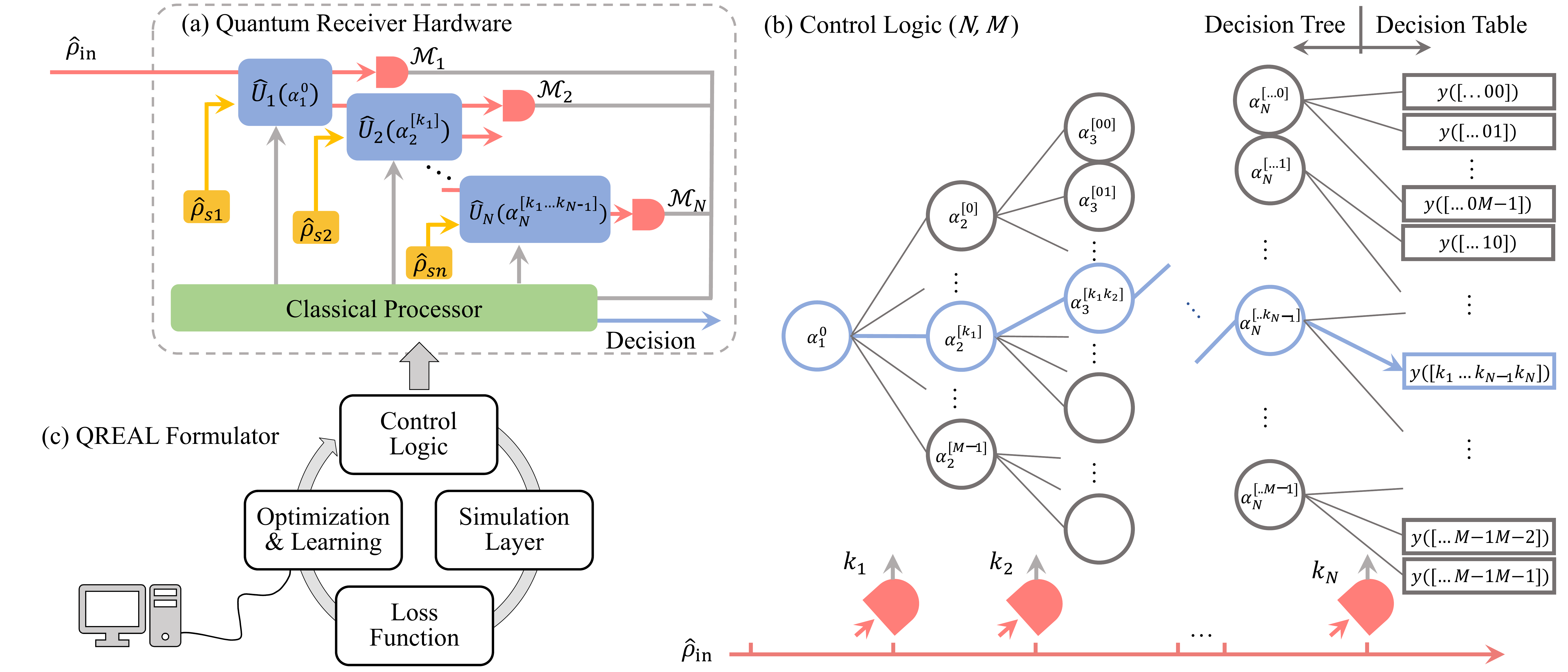}
\caption{Overview on the QREAL architecture. (a) The hardware of a general quantum receiver composed of $N$ rounds of processing. The $j$th processing round entails a variational quantum circuit represented by the unitary operation $\{\hat{U}_j\}$ configured by the parameter set $\{\alpha_j\}$, ancillary states $\rho_{s_j}$, and measurement apparatus $\mathcal{M}_j$. (b) The control logic comprised of a decision tree and a decision table. Blue lines show one possible path toward the output. Quantum-receiver output $\{y\}$ is determined by the sequence of measurement outcomes, tied to a unique path in the decision tree. (c) The QREAL formulator consisting of four modules (clockwise) as one iteration in the learning process. The control logic optimized by the QREAL formulator is compiled and built into the classical processor.}
\label{fig:Fig1} 
\end{figure*}

\textit{Introduction.}---Quantum information science (QIS) endows communication~\cite{gisin2007quantum,scarani2009security}, sensing~\cite{degen2017quantum,pirandola2018advances}, and computing~\cite{preskill2018quantum,arute2019quantum} capabilities unrivaled by their classical counterparts. QIS has also sharpened our understanding of the fundamental limits of information acquisition, transfer, and processing due to the indistinguishability of nonorthogonal quantum states, which in turn place bounds on the rate of optical communications~\cite{holevo1973bounds,caves1994quantum} and the precision of sensing~\cite{braunstein1994statistical}. Unfortunately, in many scenarios, an appreciable gap separates the performance achievable by routine measurement apparatus and what is allowed by quantum mechanics~\cite{nakahira2018optimal,PhysRevLett.127.150501}. A central theme of QIS is thus the quest for protocols that approach the ultimate performance limits.

Quantum receivers are unconventional measurement apparatuses designed to bridge the gap and enable a performance boost in a wide range of information processing tasks modeled as quantum-state discrimination~\cite{barnett2009quantum} or parameter estimation~\cite{helstrom1969quantum}. Pioneering quantum-receiver works unveiled that adaptive structures based on quantum circuits configured by feed-forward controls~\cite{dolinar1976class,assalini2011revisiting} can vastly reduce the error probability in discriminating nonorthogonal quantum states. A landmark development was the Dolinar receiver capable of discriminating two weak coherent states at the fundamental Helstrom bound~\cite{dolinar1973optimum}. More recent studies have laid out quantum-receiver structures to benefit a variety of near-term tasks encompassing quantum state tomography~\cite{andersen2015hybrid,izumi2020tomography,xu2020experimental,burenkov2020time}, target detection~\cite{zhuang2017optimum,zhuang2019physical}, communication~\cite{tsujino2011quantum,chen2012optical,becerra2013experimental,becerra2015photon,ferdinand2017multi,burenkov2018quantum,izumi2020experimental,rengaswamy2021belief}, and computing ~\cite{su2019conversion,bourassa2021blueprint}. Despite these encouraging advances, the design of quantum receivers for general QIP problems remains a formidable challenge, primarily due to the large Hilbert space that quantum states reside in. Indeed, only a few are known to achieve the ultimate performance limit even excluding imperfections. In practical situations, noise and disturbance in a dynamic environment would further bring substantial complexities to the quantum-receiver design~\cite{dimario2020phase}, rendering the traditional design method based on analytic modeling clumsy and impotent. 

The rapid development of data science has given rise to efficient tools for addressing complex data-processing problems in a large parameter space, shifting the landscape of data mining~\cite{han2011data}, computer vision~\cite{forsyth2011computer}, automated control~\cite{abbeel2007application,murphy2011littledog}, and decision making~\cite{gibney2016google}. State-of-the-art data-science methods now help encounter the unprecedented challenge of designing quantum protocols and platforms, such as engineering of quantum states~\cite{knott2016search,cui2019wave}, operators~\cite{molesky2018inverse}, or a combination of both ~\cite{arrazola2019machine,krenn2020computer}.

Here, we harness reinforcement learning~\cite{sutton2018reinforcement} to design quantum receivers, formulating the quantum receiver enhanced by adaptive learning (QREAL) architecture capable of tackling a series of quantum-state discrimination and data-classification problems. The QREAL architecture enjoys a substantially reduced complexity compared with other proposals based on model-free reinforcement learning~\cite{bilkis2020real,vargas2021quantum}, allowing it to be embodied in a photonic platform compatible with telecommunication. We then verify QREAL's capability of tackling different quantum-state discrimination problems with advantages over the standard quantum limit (SQL). By virtue of its capability of adapting to diverse operational conditions, QREAL's advantage over prior quantum receivers is further magnified in the presence of practical imperfections. These features of QREAL render it readily available to enhance long-haul communications, imaging, and sensing systems.

\begin{figure*}[htb!]
\centering
\includegraphics[width=6.5in]{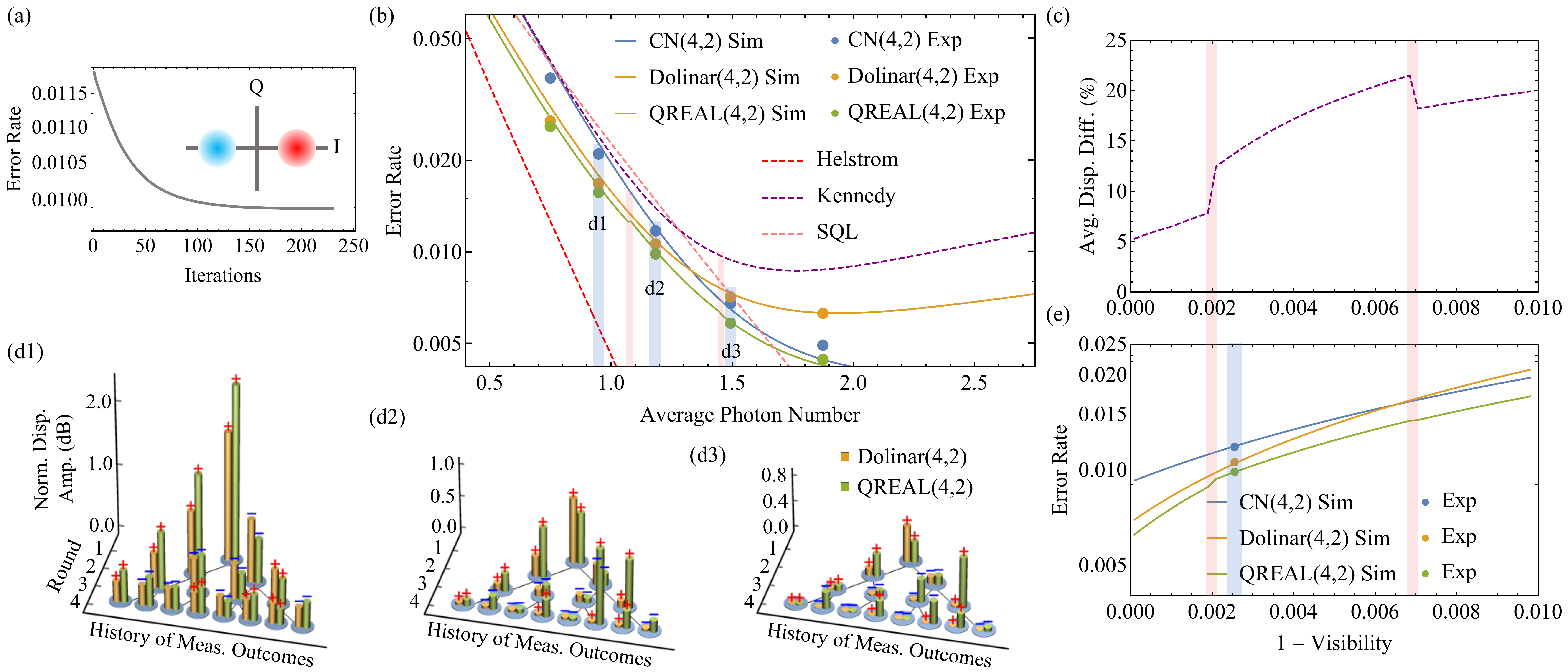}
\caption{QREAL for BPSK-encoded quantum-state discrimination. (a) Evolution of error rate during learning iterations. Inset: constellation diagram of BPSK encoding. (b) Error-rate comparison for the QREAL and other receivers in the presence of noise. An interference visibility of 99.75\% is set in simulations (solid lines) and is measured in experiments (points) for all tested receivers. The Helstrom bound and SQL (homodyne) are noise-free and loss-free. (c) Average difference in displacement amplitudes applied by the QREAL and Dolinar receiver at sub-unity visibilities, showing QREAL's ability to adapt to imperfections. The corresponding error rates are displayed in (f). Red belts mark discontinuities in quantum-receiver strategies, a sign of escaping local optimums. (d1)-(d3) show the displacement amplitudes (pillars) and phases (signs) of each node in the Dolinar(4,2) and QREAL(4,2) decision trees at the three different input photon numbers (blue belts in (b)). The displacement amplitudes are in a dB scale, normalized to these used in the CN receiver. Mean photon number in producing figure (a), (b) and (e) is $\sim$1.2.}
\label{fig:Fig2} 
\end{figure*}

\textit{QREAL Architecture.}---The general goal of our quantum-receiver is to perform a hypothesis testing task between a set of quantum states, each tagged by a classical entry $y$. A general quantum-receiver structure depicted in Fig.~\ref{fig:Fig1}(a) comprises $N$ rounds of processing and measurements. Following the control logic, the classical processor manages the quantum circuit and provides a decision based on the measurement history. In the $j$th round, the quantum state from previous steps is modified by a history-configured unitary operation $\hat{U}_j(\alpha_j^{[k_1 k_2 ... k_{j-1}]})$ with an ancillary state $\hat{\rho}_{s_j}$, then mapped into a result $k_j$ which leads to the operation in the next round.

Figure~\ref{fig:Fig1}(b) illustrates a control logic comprised of an $M$-ary decision tree with depth $N$ and a decision table, both employed by a quantum receiver specified as QREAL$(N,M)$. Each node, linked from its parent node, contains a variable that is optimized by the QREAL formulator (Fig.~\ref{fig:Fig1}(c)). The noise awareness of QREAL is accomplished by simulating a batch of decision trees, each generated with stochastically-sampled noise through all rounds. In doing so, the QREAL formulator obtains a collection of probability distributions for the measurement history conditioned on each input state, subject to the noise characteristics. Then, the decision table is derived by Bayesian inference, and the decision tree is adjusted to lower the loss function. After the iterations, an optimized control logic for a specific QIP task is loaded into the classical processor (Detailed description in Supplementary section I~\cite{supplemantary}).

\begin{figure*}[htb!]
\centering
\includegraphics[width=6.5in]{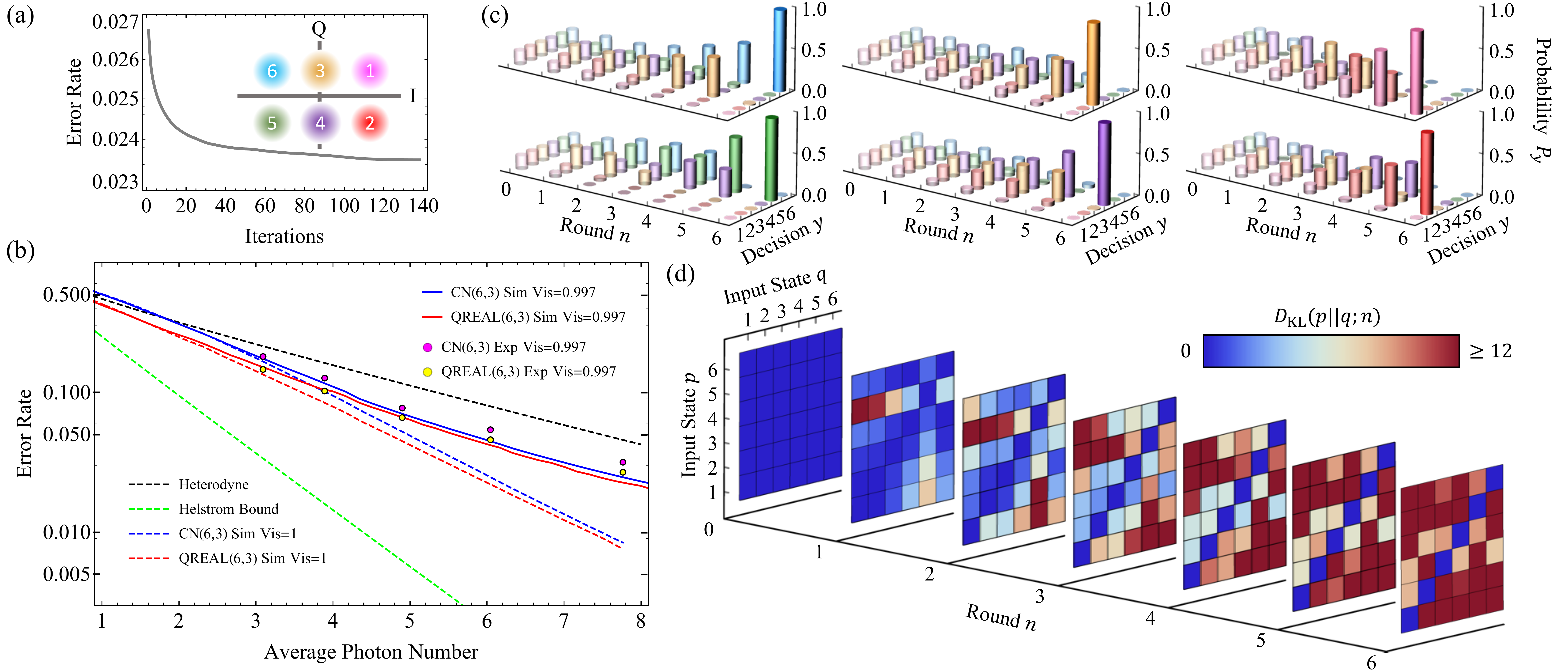}
\caption{QREAL for QAM-encoded quantum-state discrimination. (a) Evolution of error rate during learning iterations. Inset: constellation of QAM-6 encoding. (b) Comparison of the error rates for QREAL and other receivers in the presence of noise. An interference visibility of 99.7\% is set in simulations (solid lines) and is measured in experiments (points) for all tested receivers. Dashed lines are simulated error-rate lower bounds without noise. (c) Posterior probabilities for the input quantum states in different rounds. The uniform prior probabilities are plotted in Round 0. The position arrangement and color of states follows the inset in (a). (d) Evolution of the relative entropy $D_{\rm KL}(p||q;n)$ (Kullback-Leibler divergence) for the pair-wise measurement statistics conditioned on the input quantum states $|\beta_p\rangle$ and $|\beta_q\rangle$. Mean photon number used in figure (a), (c) and (d) is $\sim$7.8.}
\label{fig:Fig3} 
\end{figure*}

\textit{Experiment.}---The nonorthogonality of quantum states precludes them from being perfectly discriminated, forbidding the classical receivers operate near the fundamental limits. To bridge the gap, we leverage the QREAL formulator to construct quantum receivers for the problem of discriminating weak nonorthogonal coherent states. To date, experimentally-implemented quantum receivers are predominantly devised by analytic approaches that are incapable of adapting to diverse operational conditions including various genres of noise and imperfections, rendering their performance far inferior to that of QREAL as we show below.

The quantum-receiver hardware is a fiber-based platform operating at the c-band of optical communication. The components realize suitable functionalities within the general QREAL architecture. Specifically, practical modules of variational quantum circuits and measurements are assembled to manipulate and detect weak coherent states, involving vacuum ancillary states, a reconfigurable quadrature displacement operation and a photon-number resolving (PNR) detector. The overall fidelity of displacement operation exceeds 99.7\% (Supplementary section II \& III~\cite{supplemantary}).

As a first test, we develop and implement the QREAL robust against noise and other imperfections for the binary phase-shift keying (BPSK) format ((Fig.~\ref{fig:Fig2}(a), inset). The discrimination of quantum states encoded in the BPSK format has been extensively investigated, with many quantum receivers proposed \cite{kennedy1973near,dolinar1973optimum,dolinar1976class,assalini2011revisiting} and implemented~\cite{tsujino2011quantum} to beat the SQL. The Dolinar receiver~\cite{dolinar1973optimum,assalini2011revisiting} was the first to approach the ultimate Helstrom bound for BPSK quantum-state discrimination in an ideal scenario, but its advantage over SQL quickly fades away under noise, due to the lack of a mechanism to cope with practical nonidealities.

In contrast, QREAL(4,2) for BPSK can mitigate the adverse effect caused by noise and other imperfections by {\em adaptively} learning their patterns in $\sim$150 iterations ($\sim$15 seconds) (Fig.~\ref{fig:Fig2}(a)). The disparity between the learned displacement parameters and those used in the Dolinar(4,2) receiver is shown in Fig.~\ref{fig:Fig2}(c) and (d1)-(d3). To verify QREAL(4,2)'s robustness against noise, we test QREAL(4,2), the conditional-nulling~\cite{chen2012optical} receiver (CN(4,2)), and Dolinar(4,2)~\cite{assalini2011revisiting} subject to an identical noise pattern. The simulation results in Fig.~\ref{fig:Fig2}(b) and (e) show that, while noise quickly degrades the performance of Dolinar(4,2), QREAL(4,2) maintains its advantage over SQL within a broad range of power levels of the input quantum states. In the large photon-number or high-noise regime, QREAL picks a strategy closer to that of the CN receiver, whereas in the small photon-number or low-noise regime, the QREAL learns a strategy closer to that of the Dolinar receiver. Overall, QREAL outperforms the other two quantum receivers across the entire parameter space, shown by the experimental error rates displayed in colored dots, with error bars smaller than the size of the points. With no additional components between the displacement operation and the detector, the experimental setup achieves two parameters to beat BPSK's loss-less SQL~\cite{dimario2018robust,cook2007optical}, i.e., an overall efficiency of 85\% while maintaining the visibility at 99.75\%. Specifically, we achieve a raw bit error rate of $<2.5\%$ at a mean photon number of $\sim$0.75, corresponding to an information rate as high as 1.1 bits per received photon. QREAL enjoys an error rate 39\% below the SQL at a mean photon number of 0.95 and keeps an advantage over the SQL for mean photon numbers below 1.6. After adapting to the noise, QREAL reduces up to 14\% error rate compared with the best of the CN and Dolinar receivers~\cite{notes}.

We next develop the QREAL for the quadrature amplitude modulation (QAM) encoding (Fig.~\ref{fig:Fig3}(a), inset) to demonstrate its capability of handling complex tasks in the presence of noise. QAM profits the spectral efficiency by leveraging a larger codeword space but challenges the design of sub-SQL quantum receivers due to its larger parameter space. The QREAL formulator takes $\sim$120 iterations ($\sim$10 minutes) to converge the design at a minimal error rate (Fig.~\ref{fig:Fig3}(a)). The constructed QREAL surpasses the SQL and beats the error rate of the CN receiver. The learned QREAL(6,3) for QAM-6 decoding is tested back-to-back with CN(6,3). Both the simulation and the experimental results illustrate QREAL's noise mitigation capability, which underpins QREAL's performance gain in the low photon-number regime where noise constitutes a significant portion of error events. In practice, QREAL enables up to 43\% (average 32\%) error-rate reduction over the SQL (loss-free heterodyne) and up to 19\% (average 17\%) improvement over the CN receiver, as depicted in Fig.~\ref{fig:Fig3}(b).

To illustrate the internal workflow of QREAL, Fig.~\ref{fig:Fig3}(c) plots the evolution of the posterior probability distributions through the processing rounds for each codeword, starting from a uniform prior probability distribution. More information about the incoming quantum state is acquired as the QREAL executes through the consecutive rounds so that the correct codeword becomes prominent the posterior probability distribution while the incorrect guesses are suppressed. To further understand how the QREAL discriminates between different codewords, we quantitatively compare the distance between the measurement statistics resulted from the 6 different coherent states using the relative entropy $D_{\rm KL}(p||q)$, as shown in Fig.~\ref{fig:Fig3}(d). By the end of the first two rounds, the measurement statistics for a subset of the codeword states, e.g., $|\beta_1\rangle$ and $|\beta_5\rangle$, have been adequately disparate so that they are distinguished with confidence. The rest of the codeword states, however, remain unsure due to their close measurement statistics. As the processing proceeds into subsequent rounds, more measurement outcomes lead to distinct measurement statistics for different codeword states, allowing QREAL to pick the correct codeword. 

\textit{Discussion.}---The quantum-receiver hardware can be augmented with new types of variational quantum circuits and measurements to embrace a wider scope of QIP problems. For example, continuous-variables measurements such as the homodyne detector offers an additional degree of freedom to access a richer set of information~\cite{xu2020experimental}. With a versatile topology, the QREAL formulator can be generalized to most near-term measurement-based quantum applications, including quantum state generation~\cite{su2019conversion}, tomography~\cite{izumi2020tomography}, variational quantum eigensolvers~\cite{cerezo2021variational}, and quantum-enhanced sensing~\cite{zhuang2019physical}, after proper discretizations.

Moving toward practical operational environments, the QREAL formulator can fit the \textit{in-situ} learning iterations as tests on the hardware also provide the sampled probability distribution. In doing so, the QREAL formulator will be more efficiently in dealing with the exponentially large quantum-state space and, in the meantime, adapt to slowly-varying noise. This feature enables long-time unsupervised stability, which is crucial for real-world applications. In addition, the QREAL formulator supports artificial neural networks in its decision strategy, for capturing patterns embedded in data~\cite{sentis2019unsupervised}, like the variational autoencoder~\cite{pu2016variational}. In circumstances without prior knowledge for the channel and receiver imperfections, model-free reinforcement-learning techniques can be used to design quantum receivers~\cite{bilkis2020real,vargas2021quantum}.

A few remarks on the optimization process of QREAL are worth making. A recent paper discovered that the initial parameters, to a large extent, determines the local minimum a machine-learning algorithm converges to~\cite{PhysRevLett.127.120502}. In our work, the QREAL formulator leverages a greedy algorithm to generate initial parameters that obey a preset criterion, locating near those of the CN receiver. Other sets of initial parameters may exist to further enhance the performance of QREAL, but as a tradeoff they would require more computing resources to identify. In addition, the initialization processes for QREALs tackling more complex quantum information processing problems would call for more systematic studies.

\textit{Conclusions.}---We have developed QREAL to address two quantum-state discrimination problems. First, QREAL for BPSK is shown to outperform both the Dolinar and the CN receivers. Notably, QREALs are proven robust against imperfections so that QREAL's performance advantage over conventional quantum and classical receivers sustains over a large noise region. We also constructed QREAL for QAM to verify its capability of undertaking a complex QIP task that require optimization in a large parameter space with more than 1000 variables. The QREAL architecture is envisioned to endow new functionalities in the noisy intermediate-scale quantum era by harnessing the hybrid quantum-classical information processing architecture.

\textit{Acknowledgements.}---We gratefully acknowledge funding support by the National Science Foundation Grants No.~CCF-1907918, No.~ECCS-1828132, and No.~EEC-1941583. Z.Z. acknowledges NSF CAREER Award No.~ECCS-2144057. Q.Z. acknowledges support from Defense Advanced Research Projects Agency (DARPA) under Young Faculty Award (YFA) Grant No. N660012014029. C.C. thanks Kaushik Seshadreesan and Masoud Babaeian for their help.


\begin{thebibliography}{54}
\bibitem[{sup()}]{notes}
\bibinfo{note}{Averaging over the five tested mean photon numbers, QREAL's error rate is 13\% lower than the Dolinar receiver, 19\% lower than the CN receiver, and 17\% lower than the SQL. Averaging over four tested mean photon numbers below 1.5, QREAL's error rate is 32\% lower than the SQL. By incorporating photon-number-resolving detectors [49], QREAL will be able to surpass the SQL over a wider range of mean photon numbers.}
\end{thebibliography}



\section*{Supplementary material}

# Supplemental Material: Quantum Receiver Enhanced by Adaptive Learning

Chaohan Cui,[1] William Horrocks,[1] Shuhong Hao,[2] Saikat Guha,[1,3]
N. Peyghambarian,[1,2] Quntao Zhuang,[3,1] and Zheshen Zhang[2,3,1,*]

[1]*James C. Wyant College of Optical Sciences,
The University of Arizona, Tucson, Arizona 85721, USA*
[2]*Department of Materials Science and Engineering,
The University of Arizona, Tucson, Arizona 85721, USA*
[3]*Department of Electrical and Computer Engineering,
The University of Arizona, Tucson, Arizona 85721, USA*

## I. DETAILED MODEL OF THE QREAL ARCHITECTURE

### Hardware

In Fig. 1(a), The $j$th round entails a variational quantum circuit described by a reconfigurable unitary operation $\hat{U}_j(\alpha_j^{[k_1 k_2 \ldots k_{j-1}]})$, an ancillary quantum state $\hat{\rho}_{s_j}$, and a quantum measurement $\mathcal{M}_j$. To start the quantum-receiver operations, the first variational quantum circuit $\hat{U}_1(\alpha_1^0)$ ($[x_0] = 0$) takes the input in state $\hat{\rho}_{\rm in}$ and the ancilla in state $\hat{\rho}_{s_1}$ and diverts a portion of its output quantum state to the measurement device. A control logic running on a classical processor then generates, in real time, a feed-forward signal $\alpha_2^{[k_1]}$ based on the measurement data $k_1$ so that the second variational quantum circuit is configured to $\hat{U}_2\left(\alpha_2^{[k_1]}\right)$. After $N$ rounds of processing and measurements the classical processor selects an entry $y$ from a decision table as the output of the quantum receiver. The control logic hence defines the quantum-receiver strategy.

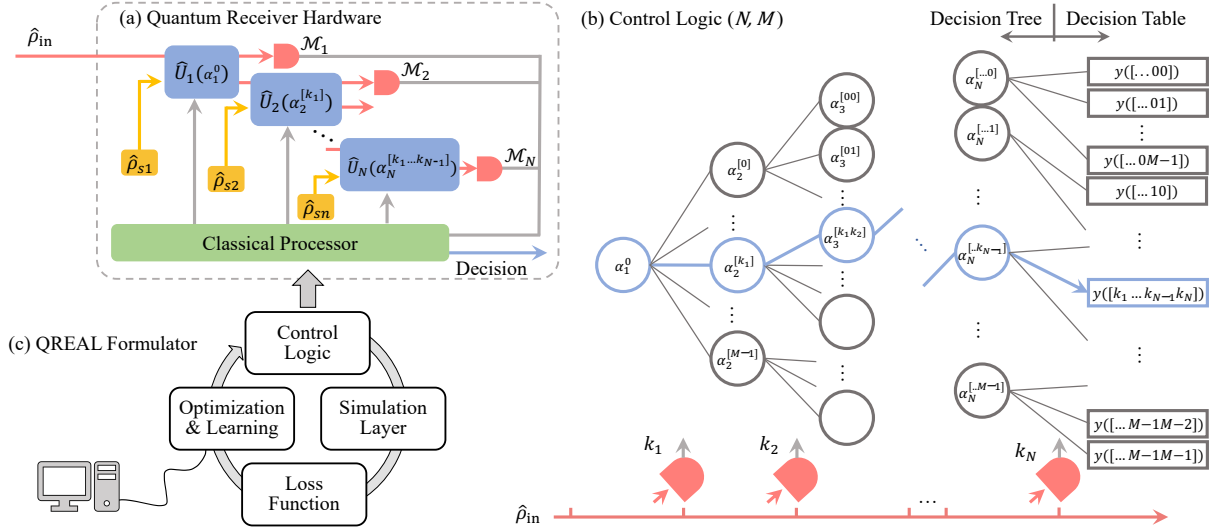

FIG. 1. A copy of Fig. 1 from the main paper: Overview on the QREAL architecture. (a) The hardware of a general quantum receiver composed of $N$ rounds of processing. The $j$th processing round entails a variational quantum circuit represented by the unitary operation $\{\hat{U}_j\}$ configured by the parameter set $\{\alpha_j\}$, ancillary states $\rho_{s_j}$, and measurement apparatus $\mathcal{M}_j$. (b) The control logic comprised of a decision tree and a decision table. Blue lines show one possible path toward the output. Quantum-receiver output $\{y\}$ is determined by the sequence of measurement outcomes, tied to a unique path in the decision tree. (c) The QREAL formulator consisting of four modules (clockwise) as one iteration in the learning process. The control logic optimized by the QREAL formulator is compiled and built into the classical processor.

## Control logic

Figure 1(b) illustrates a control logic comprised of an $M$-ary decision tree with depth $N$ and a decision table, both employed by a quantum receiver specified as QREAL$(N, M)$. A node at the $j$th layer of the control logic carries a parameter $\alpha_j^{[x_{j-1}]}$ that specifies the setting of the variational quantum circuit in the $j$th round according to the measurement outcomes of previous $j-1$ rounds $[x_{j-1}]$. The control logic starts out at the root node ($j=1$) and hops to a node in the next layer contingent upon the measurement outcome $k_1$, deriving the history records $[x_1] = [k_1]$. The migration in the decision tree continues until landing in an entry $[x_N] = [k_1...k_N]$ with $y([x_N]) \in \mathcal{D}$ in the decision table as the output of QREAL, where $\mathcal{D}$ is a set of possible output values. An exemplar control-logic path is highlighted by the blue trajectory in Fig. 1(b).

## Formulator

The QREAL formulator drawn in Fig. 1(c) aims to optimize the control logic for a specified QIP problem. The formulator iterates between a simulation layer, a loss function, and the optimization and learning algorithm to update the control logic. The simulation layer processes on the input density matrix and generates the measurement statistics characterized by the posterior probability distributions sampled during the execution of the quantum receiver. Deviations caused by noise and imperfections are also sampled through the simulation. The loss function is then derived from the measurement statistics to quantify the goal of optimization, such as the error rate. The classical optimization and learning algorithm subsequently updates the control logic based on the loss-function value and the gradient of its variables.

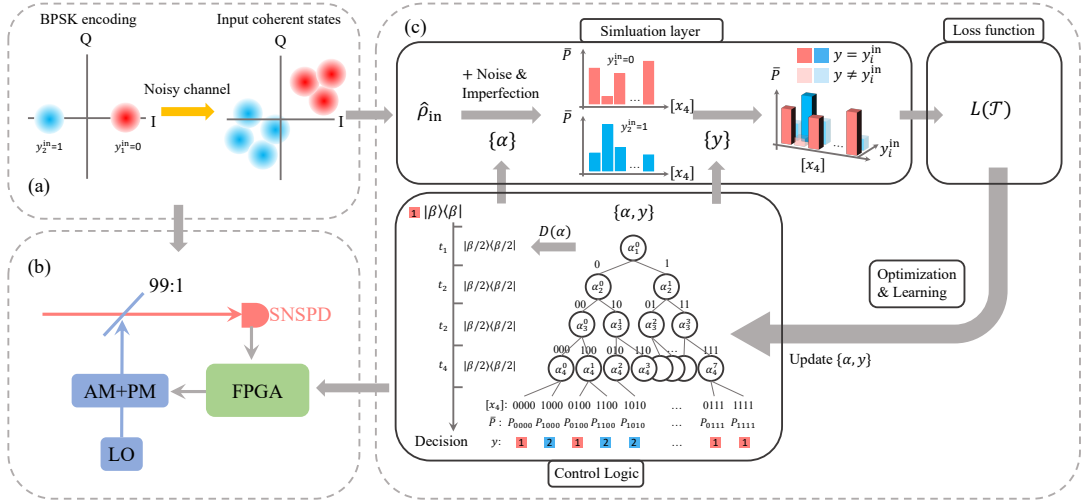

FIG. 2. QREAL(4,2) for BPSK. (a) Quantum-state evolution through a noisy channel. (b) The quantum-receiver for weak coherent-state discrimination. See Fig.3(a) for more details. (c) The formulator for QREAL(4,2). $[x_4]$ is represented in decimal numbers. $\{\alpha_j^{[x_j]}\}$ is the parameter set for displacements at the $j$th round. $\{[x_4]\}$ is the history of measurement outcomes. $\{\bar{P}\}$ is the probability distribution associated to input state $|\beta\rangle\langle\beta|$ and the history $[x_4]$. $\{y\}$ is the decision based on $\{[x_4]\}$.

## II. THEORY OF THE QREAL FOR WEAK COHERENT-STATE DISCRIMINATION

The objective of weak coherent-state discrimination is to minimize the error probability in labeling a quantum state belonging to the codeword set $\{|\beta_j\rangle\langle\beta_j|\}_{j=1}^{L}$ of $L$ coherent states. A general description for the coherent-state ensemble is the density matrix $\hat{\rho}_{\text{in}} = \sum_{i=1}^{L} P_i^{\text{in}} |\beta_i\rangle\langle\beta_i|$, where $P_i^{\text{in}}$'s are the prior probabilities. In the main text, the density matrix for the BPSK encoding with equal prior probabilities is $\hat{\rho}_{\text{in}} = \frac{1}{2}|\beta\rangle\langle\beta| + \frac{1}{2}|-\beta\rangle\langle-\beta|$, whereas the density matrix for the QAM-6 encoding with equal prior probabilities is $\hat{\rho}_{\text{in}} = \sum_{i=1}^{6} \frac{1}{6}|\beta_i\rangle\langle\beta_i|$, $\beta_i \in \{\beta e^{i\tan^{-1} 1/2}, \beta e^{-i\tan^{-1} 1/2}, i\beta/\sqrt{5}, -i\beta/\sqrt{5}, -\beta e^{i\tan^{-1} 1/2}, -\beta e^{-i\tan^{-1} 1/2}\}$. The main photon number of BPSK is $|\beta|^2$ and of QAM-6 is $\frac{11}{15}|\beta|^2$ in front of the receiver. Each encoding picks a pure state $|\beta_i\rangle\langle\beta_i|$ from the codeword state set with the label $y_i^{\text{in}} \in \boldsymbol{y}^{\text{in}}$. A noisy channel contaminates the pure state, resulting a mixed state $\rho_i$ at the quantum receiver.

The variational quantum circuit for the $j$th round of the quantum-receiver hardware sketched in Fig. 2(b) consists of a beamsplitter tapping $1/N$ portion of the input quantum state $\hat{\rho}_{\text{in}} = |\beta\rangle\langle\beta|$ followed by a reconfigurable quadrature displacement operation $\hat{U}_j(\alpha_j) = \exp(\alpha_j \hat{a}^{\dagger} - \alpha_j^* \hat{a})$ on the tapped coherent state $|\beta/\sqrt{N}\rangle$. A photon-number resolving (PNR) detector then measures the displaced state $|\beta/\sqrt{N} + \alpha_j\rangle$, yielding $k_j$ counted photons as its output. The classical processor subsequently collects the output from time-multiplexed PNR detector and sets the amount of displacement applied in the variational quantum circuit of the next round.

Suppose that the duration of the input quantum state is $T$ seconds, the depth of control logic is $N$, and the PNR can resolve up to $M-1$ photons. The $j$th round is executed over $t_j$ seconds so that $T = \sum_{j=1}^{N} t_j$. The measurement outcome $k_j \in \{0, 1, ..., M-2, \geq M-1\}$ determines the route to the next node in the decision tree. The measurement outcomes in the first $j$ rounds is assembled to form the sequence $[x_j] \equiv [k_1 k_2 ... k_{j-1} k_j]$, $j \leq N$.

The effect of noise and other imperfections is modeled by an additive probability $f_N$, which depends on the input coherent state $|\beta_i\rangle\langle\beta_i|$, the displacement $\alpha_j^{[x]}$ applied in the $j$th round, the duration of the $j$th round $t_j$, and the measurement outcome $k_j$. During the learning process, the noise term $f_N$ is sampled independently and added to a batch of decision trees, without resorting to any analytical form. With the input coherent state $|\beta_i\rangle\langle\beta_i|$ labeled by $y_i^{\text{in}}$, the conditional probability of attaining $k_j$ in PNR measurement reads

$$P_j(k_j \mid y_i^{\text{in}}, \alpha_j^{[x_{j-1}]}, f_N) = \begin{cases} |\beta_i \sqrt{t_j/T} + \alpha_j^{[x_{j-1}]}|^{2k_j}/k_j! + f_N\left(\alpha_j^{[x_{j-1}]}, \beta_i, t_j, k_j\right), & k_j \in \{0, 1, ..., M-2\} \\ 1 - \sum_{k'=0}^{k'<M-1} P_j\left(k' \mid \alpha_j^{[x_{j-1}]}, y_i^{\text{in}}, f_N\right), & k_j = M-1 \end{cases}. \tag{1}$$

The size-$D$ batch-averaged probability of recording a specific sequence of $[x_N]$ conditioned on the input-state label $y_i^{\text{in}}$, sampled noise function $f_N \in \mathcal{D}$, and the displacement settings of all $N$ rounds, defined as $\boldsymbol{\alpha}_N^{[x_{N-1}]} \equiv \{\alpha_1^0, \alpha_2^{[x_1]}, \alpha_3^{[x_2]}, ..., \alpha_N^{[x_{N-1}]}\}$, along the decision path set by $[x_N]$ is thus

$$\bar{P}\left([x_N] \mid y_i^{\text{in}}, \boldsymbol{\alpha}_N^{[x_{N-1}]}\right) = \frac{1}{D} \sum_{f_N \in \mathcal{D}} P_1(k_1 \mid \alpha_1^0, y_i^{\text{in}}, f_N) \prod_{j=2}^{N} P_j\left(k_j \mid y_i^{\text{in}}, \alpha_j^{[x_{j-1}]}, f_N\right). \tag{2}$$

The output of QREAL conditioned on the sequence of measurement outcomes is given by the label of the input coherent state that maximizes the conditional probability of registering this sequence:

$$y\left([x_N] \mid \boldsymbol{\alpha}_N^{[x_{N-1}]}\right) = \arg\max_{y_i^{\text{in}}} P_i^{\text{in}} \bar{P}\left([x_N] \mid y_i^{\text{in}}, \boldsymbol{\alpha}_N^{[x_{N-1}]}\right) \tag{3}$$

The loss function $L$ in turn is defined as the overall error rate given the entire decision tree $\mathcal{T}$, defined as

$\mathcal{T} \equiv \left\{ \boldsymbol{\alpha}_N^{[x_N-1]} \right\}$, over all $[x_N]$.

$$L(\mathcal{T}) = \sum_{[x_N],i} \left(1 - \delta_{y_i^{\text{in}}, y\left([x_N] | \boldsymbol{\alpha}_N^{[x_N-1]}\right)}\right) P_i^{\text{in}} \bar{P}\left([x_N] \mid y_i^{\text{in}}, \boldsymbol{\alpha}_N^{[x_N-1]}\right), \tag{4}$$

where $P_i^{\text{in}}$ is the prior probabilities for each input quantum state, and the Kronecker delta function $\delta_{y_i^{\text{in}}, y}$ ensures that only the possibilities for incorrect decisions that fail to identify the input quantum-state label contributes to the loss function.

The above algebras fulfill the requirement of automatic differentiation needed by backpropagation, which is exploited in the chosen machine-learning algorithm to optimize QREAL's control logic by gradient until landing in a local minimum of the loss function.

The machine-learning algorithm leverages stochastic gradient descent with adaptive momentum. The initial control logic and decision table are both picked by a greedy strategy. This greedy strategy employs stochastic optimization over the displacement phases while fixing the magnitude so that it tends to nullify the most-likely quantum state. The initial control logic and decision table are identical to these of the CN receiver in the absence of phase noise, but start to deviate from these of the CN receiver when phase noise becomes prominent.

A complete illustration for the QREAL(4,2) formulator for BPSK with equal $t_j$'s is drawn in Fig. 2(c).

The relative entropy between measurement statistics associated with two incoming quantum-state labels $y_p^{\text{in}}$ and $y_q^{\text{in}}$ upon the completion of the $n$ round is estimated by

$$D_{\text{KL}}(p||q;n) = \sum_{[x_n]} \bar{P}\left([x_n] \mid y_p^{\text{in}}, \boldsymbol{\alpha}_n^{[x_n-1]}\right) \log_2 \left( \frac{\bar{P}\left([x_n] \mid y_p^{\text{in}}, \boldsymbol{\alpha}_n^{[x_n-1]}\right)}{\bar{P}\left([x_n] \mid y_q^{\text{in}}, \boldsymbol{\alpha}_n^{[x_n-1]}\right)} \right), \quad \text{with } p, q, n \in \{1, 2, 3, 4, 5, 6\}. \tag{5}$$

A larger relative entropy renders two quantum states more distinguishable, thereby leading to less ambiguity and lower error rates.

## III. EXPERIMENTAL SETUP AND CALIBRATION

The experiment entails the QREAL formulator and the quantum-receiver hardware that routinely communicate via an ethernet cable. The QREAL formulator running on a desktop computer continuously learn and deliver the optimized control logic to the classical processor of the quantum-receiver hardware. The classical processor consisting of an FPGA operates the control logic and collects the data produced by the quantum-receiver hardware. To initialize the QREAL formulator, the FPGA runs an estimation procedure to infer the average number of received photons, fringe visibility, and other characteristics of QREAL.

### The QREAL formulator

The QREAL formulator is programmed in Python using TensorFlow, an open-source machine-learning library that supports parallel computing, symbolic graph computing, and automatic differentiation. The noise and device imperfections are simulated by a batch of decision trees with stochastically-sampled dark counts, phase-locking noise, and amplitude-modulation noise. The employed batch sizes are 1000 for BPSK and 100 for QAM-6.

Prior to the iterative learning procedure, the QREAL formulator first uses a greedy algorithm to create an initial control logic composed of a decision tree that carries the displacements applied by the variational quantum circuits and a decision table that stores the outputs of the QREAL. During the iterations, the

simulation layer simulates the evolution of batches of input quantum states through the variational quantum circuits configured by the current control logic with sampled noise and derives the measurement outcomes in all rounds using the coherent-state discrimination model (Appendix II), based on which the posterior probability distribution for each input quantum state from the codeword set is formed. Each entry in the decision table ties to a distinct sequence of measurement outcomes and labels the input quantum state that yields the largest posterior probability given that sequence. Next, the loss function and its gradients to the control-logic parameters are calculated based on the simulated probability that QREAL's output differs from the input quantum-state label, i.e., the error rate. Finally, the optimization and learning module updates the control logic by adjusting its parameters toward a reduced error rate. The updated control logic will be used in the simulation layer in the next iteration until the loss function converges.

## Hardware platform

The variation quantum circuit is a displacement operation implemented by an interference between the quantum-signal light and a phase-stabilized strong local oscillator (LO) on a 99:1 beamsplitter, as sketched in Fig. 2(b). The amplitude and phase modulators in the LO arm are controlled by a field-programmable gate array (FPGA), which processes the measurement data from a superconducting nanowire single-photon detector (SNSPD) to follow the QREAL control logic. Achieving a high overall efficiency of the circuit is critical to beat standard quantum limit (SQL). We remove all lossy components between the displacement operation and the detector, resulting in the highest efficiency (∼85%) reported to date. Meanwhile, the system achieves high visibility (>99.7%) without trading its efficiency by exploiting back-propagation locking.

The quantum-receiver hardware platform illustrated in Fig. 3(a) operates at 1550 nm for compatibility with fiber-optic networks. Due to the required stable relative phase between the LO and and quantum-signal light, the system switches periodically between a phase-locking stage and the data-acquisition stage, each consuming 100 μs, as illustrated in (Fig. 3(b)). In the phase-locking stage, a shutter is opened to inject

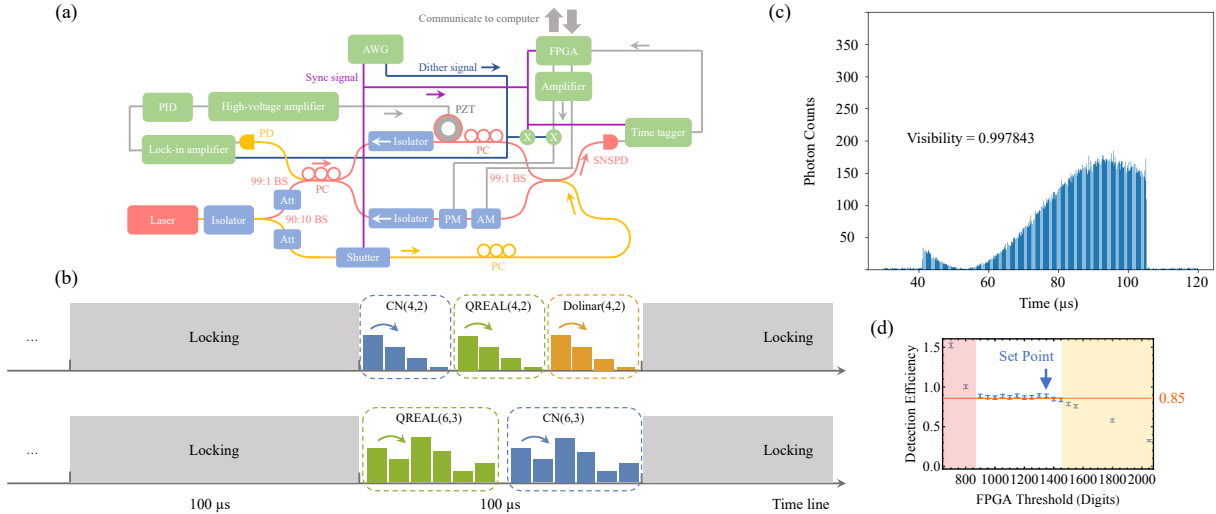

FIG. 3. (a) Experimental setup. The quantum-receiver hardware platform includes a Mach-Zehnder interferometer (red), a phase-locking module (yellow), fiber-optic components (light blue), electronic devices (green), electrical synchronization signals (magenta), the dither (dark blue), and control signals (gray). (b) The timelines of phase-locking and data-aquisition stages for (4,2) receivers and (6,3) receivers. All quantum-receiver schemes are tested in a time-multiplexed sequence subject to nearly the same noise condition. (c) The histogram of interference visibility recorded in 5s of a 100 μs span by linearly scanning PM in the phase-locking state. The histogram is a screenshot of the hardware monitor. (d) Detector efficiency vs FPGA threshold. Threshold set point is chosen after calibration.

5

counter-clockwise locking light (orange) into the Mach-Zehnder interferometer. A dither loop stabilizes the phase difference between the LO and the quantum-signal light. The dither produced by a lock-in amplifier introduces phase modulation at 31.75 kHz. The same lock-in amplifier demodulates the output from the photo diode detecting the locking light to generate the error signal, which is subsequently processed by a PID and amplified by a high-voltage amplifier to drive a piezoelectric fiber stretcher that compensates for the phase difference. In the data-acquisition stage, the counter-clockwise locking light and the electronic dither are both blocked, the quantum-signal light is launched, controlled by the FPGA, detected by an SNSPD followed by a time tagger. A 5 kHz signal synchronizes the FPGA, the shutter, the modulators, and the time tagger.

The phase-locking fluctuation, the ambient light, the initial calibration errors, and the SNSPD dark counts introduces noise to the system. We calibrate the mean photon number, noise, and other imperfections by tracking the photon histograms while scanning the LO phase and amplitude, which in turn inform the initialization and noise samples for the QREAL formulator. The construction of the histogram is triggered by the synchronization signal. Each histogram is accumulated over 5 seconds. Our setup maintains a visibility over 99.75%, shown in a screenshot of the histogram monitor in Fig. 3(c).

The detection efficiency is dependent on the FPGA threshold set to read the electronic pulses from SNSPD. The connection between the FPGA threshold and detector efficiency is exhibited in Fig. 3(d). Low thresholds in the red shade result in overcounting the SPSPD pulses. In contrast, high thresholds in the orange shade undercounter the SNSPD pulses, thereby reducing the overall efficiency. By appropriately choosing the FPGA threshold, the overall efficiency is calibrated to be $\sim 85\%$. The PNR detection is realized by time-multiplexed on-off detection in a 5 $\mu$s windows, which is much longer than the 25 ns detector deadtime.

In the experiment, the BPSK format is tested for a continuous duration of 800 seconds, recording 4 million samples of each design at a given mean photon number. QAM-6 is tested for a continuous duration of 1200 seconds, obtaining 6 million samples for each QREAL design at a given mean photon number. All the data are first recorded in the FPGA and then analyzed offline.

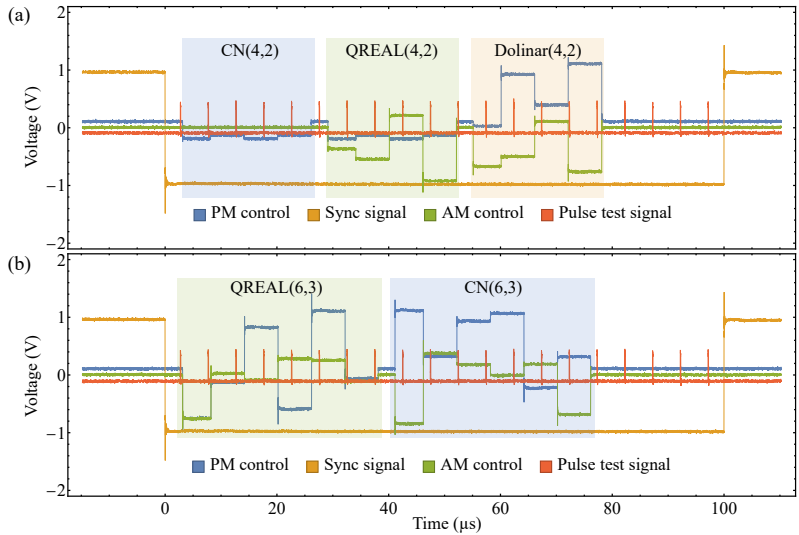

FIG. 4. (a) A sample of receiver(4,2) FPGA runtime patterns during logic tests. (b) A sample of receiver(6,3) FPGA runtime patterns during logic tests. PM control voltage and pulse test signal are shifted up and down, respectively, for better visualization. By setting the inter-bin buffer to around 1 $\mu s$ (125 clock cycles), ringing artifacts are avoided, and the search through LUT is guaranteed. There is also an inter-frame buffer between different receiver designs for switching LUTs and recording results. The pulse test signal is produced by the second FPGA and triggered by the sync signal. The test pulse repetition rate is a runtime variable for traversing all possible routines of quantum receivers.

6

## IV. FPGA CONFIGURATIONS AND CHARACTERISTICS

We choose the STEMlab 125-14 RedPitaya board to develop the QREAL. To keep the latency at a minimum, all necessary data (ML optimized voltages and count values) are stored in the programmable logic (PL) block RAM. Approximately 7% of the available 36 kbit block RAM tiles (4 of the total 60) are used by the lookup table (LUT) entries and count data. Given the low block RAM usage, more memory-intensive applications can be readily implemented using our PL design. Interactions with the block RAM are achieved using C programs, custom memory read/write cores, and combinations of basic tasks, such as updating the LUT, recording data for a user-entered amount of time, and writing the count data to a designated file, are automated with shell scripts. Parameters such as bin size, inter-bin buffers, inter-frame buffer, detection threshold, recording time are runtime configurable via GPIOs. All custom bitstreams are developed in Verilog by Vivado v2020.1 using open-source projects as templates and augmented using custom cores to achieve our desired performance.

Components of the QREAL design are verified in Vivado's simulation environment, in field tests, or a combination of both. The SNSPD output is characterized using the RedPitaya board, and multiple schemes are tested to guarantee accurate pulse detection. After implementing and testing multiple detection algorithms in the field, the threshold-based decision logic is chosen. A pulse is detected if the input voltage is below the user entered threshold on one clock cycle and above it the next. The threshold value is experimentally chosen and tuned for our setup. With an attenuated laser connected to the SNSPD, the average photon count per second is seen to be in reasonable agreement with the commercial time tagger measurement. After characterizing the detection response of the SNSPD, we use a separate FPGA to simulate the trigger signal and a custom SNSPD response. The period and amplitude of both outputs, along with pulse width, duration, and detection pattern, are configurable at runtime, shown in Fig. 4. The above feature enables us to verify the logic flow of each type of receiver before implementing it to the quantum-receiver hardware platform.

## V. QREAL'S PERFORMANCE VS. PREVIOUS SETUPS FOR BPSK

The QREAL's performance is benchmarked with previous demonstrations of decoding BPSK.

|  | Efficiency | MPN range to beat SQL | ER @ MPN=1[†] | Detection | $\lambda$ |
| --- | --- | --- | --- | --- | --- |
| Cook, Martin, & Geremia [1] | 35% | <0.25 | 12% | On/off | 852 nm |
| DiMario & Becerra [2] | 72% | < 0.5 | 3% | PNR | 633 nm |
| QREAL(4,2) (This work) | 85% | < 1.6 * | 1.6% | On/off * | 1550 nm |

TABLE I. Comparison between state-of-the-art experimental results for quantum receivers for BPSK encoding. The "ER @ MPN = 1" column shows the lowest experimental error rates at mean photon number 1. Fringe visibility of more than 99.7% achieved in all platforms. MPN: mean-photon-number; ER: error rate. (†) Experimental error rate at a mean photon number around one. (*) The introduction of PNR detectors can extend the mean-photon-number range by several folds according to simulations in Ref. [2].

---



\end{document}